\documentclass{article}
\usepackage{graphicx} 
\usepackage{hyperref}
\usepackage{caption}
\usepackage{subcaption}
\usepackage{amsmath}
\usepackage[margin=1in]{geometry}
\usepackage{setspace}
\usepackage{authblk}
\usepackage{xcolor}
\usepackage[normalem]{ulem}
\usepackage{bm}

\title{Edge Dislocation Mediated Anomalous Charge Transfer in Face Centered Cubic High Entropy Alloys\footnote{Notice: This manuscript has been coauthored by UT-Battelle, LLC, under Contract No. DE-AC0500OR22725 with the U.S. Department of Energy. The United States Government retains and the publisher, by accepting the article for publication, acknowledges that the United States Government retains a non-exclusive, paid-up, irrevocable, world-wide license to publish or reproduce the published form of this manuscript, or allow others to do so, for the United States Government purposes. The Department of Energy will provide public access to these results of federally sponsored research in accordance with the DOE Public Access Plan (\href{http://energy.gov/downloads/doe-public-access-plan}{http://energy.gov/downloads/doe-public-access-plan}).}}

\author[1]{Gautam Anand\thanks{gautamanand.mst@itbhu.ac.in}} 
\author[2]{Swarnava Ghosh\thanks{ghoshs@ornl.gov}} 
\author[3]{Suman Chabri\thanks{2021mmp001.suman@students.iiests.ac.in}}
\author[2]{Markus Eisenbach\thanks{eisenbachm@ornl.gov}} 
\affil[1]{School of Materials Science and Technology, IIT-BHU, Varanasi, UP, India 201005}
\affil[2]{National Centre for Computational Sciences, Oak Ridge National Laboratory, 1 Bethel Valley Rd, Oak Ridge, 37830, TN, USA}
\affil[3]{Department of Metallurgy and Materials Engineering, IIEST Shibpur, Howrah, WB, India 711103}

\date{}
\begin{document}

\maketitle
\begin{abstract}
Charge transfer in concentrated alloys governs their structural stability and functional response, and can be strongly perturbed by lattice defects. In high-entropy alloys, the interaction between edge dislocations and atomic volume misfit plays a central role in solid-solution strengthening models; however, the influence of dislocations on local charge transfer has not been explicitly investigated. In this work, large-scale \emph{ab initio} calculations are employed to examine dislocation-mediated charge transfer in CoNi, CoCrNi, and CoCrFeMnNi alloys. The calculations reveal anomalous charge redistribution near edge dislocation cores, including deviations from conventional electronegativity trends. The observed behavior is shown to originate from collective electronegativity equalization effects rather than simple pairwise atomic interactions. Furthermore, the asymmetric atomic-volume response within the compressive and tensile regions of the dislocation field is rationalized in terms of anomalous magneto-volume fluctuations. These results establish a direct coupling between dislocation-induced electronic redistribution and local volumetric response in chemically complex alloys. The demonstrated coupling between dislocation-mediated charge transfer and atomic volume fluctuations provides a pathway toward electronically informed solid-solution strengthening models and defect-aware alloy design strategies for chemically complex alloys. These findings further suggest that local electronic redistribution near dislocation cores can play a critical role in governing deformation behavior and defect energetics in high-entropy alloys.     
\end{abstract}

\section{Introduction}

Charge transfer between constituent atoms is a central concept in understanding bonding, phase stability, and functional properties of alloys. In classical alloy formation theory, electronic charge redistribution is typically assumed to follow simple trends governed by elemental electronegativity, work function, or valence electron concentration \cite{senkov2016new}. Within this picture, electronic charge flows monotonically from electropositive to electronegative species, and atomic charge states are regarded as intrinsic elemental characteristics weakly perturbed by alloying \cite{pettifor1987quantum}. Such assumptions form the basis of many phenomenological models describing alloy formation \cite{miedema1992energy}, chemical ordering \cite {ducastelle1993order}, and defect energetics \cite{zhang2019dissipation}.

The charge transfer influences the atomic-scale distortion in alloys dictated by the local atomic coordination environment \cite{tong2020severe,meng2021charge}, and element-resolved local atomic distortion \cite{oh2021element}. Furthermore, the atomic level stresses \cite{odbadrakh2019electronic,kohyama2021ab,ghosh2024violation}, lattice parameter of alloys \cite{tandoc2025bond}, the mechanical properties of alloys \cite{tian2025first,su2024fluctuations,shang2025temperature}, vacancy formation energy  \cite{linton2025mechanistic}, catalytic properties \cite{li2025electron}, amorphization tendencies in high-entropy alloys \cite{bu2024elastic} are also influenced by charge transfer. However, growing experimental evidence and advances in first-principles electronic structure calculations have revealed that charge transfer in high-entropy materials (HEMs) does not consistently follow traditional charge-transfer principles \cite{osinger2024charge}.

High-entropy materials (HEMs) represent a distinct class of chemically disordered crystalline solids characterized by multi-principal components in near-equiatomic or concentrated proportions, frequently stabilizing into single-phase configurations. In metallic systems, high-entropy alloys (HEAs) preferentially adopt disordered solid solutions \cite{cantor2004,HEA_Yeh2004,george_high-entropy_2019,george:2020,varvenne:2017}, whereas in ceramic systems, this paradigm has been extended to high-entropy oxides (HEOs) through the random distribution of multiple distinct cations across specific sublattices \cite{sarkar2019high,aamlid2023understanding,sun2021high,anand2018phase}. Thermodynamic stability in these systems is conventionally attributed to elevated configurational entropy of mixing, which lowers the Gibbs free energy sufficiently to counter enthalpic driving forces for phase separation, thereby suppressing intermetallic ordering, multi-phase precipitation, or local phase segregation \cite{HEA_Yeh2004,anand2016role}. Unlike many conventional alloys, which readily undergo precipitation under stress or thermal exposure \cite{ghosh2020influence,ghosh2021precipitation}, several HEAs exhibit remarkably stable solid-solution phases. Their chemically complex local environments give rise to substantial variations in vacancy formation and diffusion energetics \cite{ponga2022effects,Fani}. Since their inception~\cite{HEA_Yeh2004,cantor2004}, and the subsequent expansion into non-metallic systems, these vast compositional spaces have drawn significant attention due to their exceptional, highly tunable properties. While HEAs offer distinct mechanical and structural profiles, including superior yield strength to ductility combinations \cite{li2022strain,zhong2023deciphering,jian2020effects,gupta2022deformation}, radiation damage resistance \cite{orhan2023electronic} and enhanced corrosion resistance~\cite{HEA_Yeh2004,cantor2004,HEA_SC,li_mechanical_2019,george_high-entropy_2019,zhou2023,liu2019dislocation}, HEOs introduce novel functional degrees of freedom, exhibiting tailored electronic, catalytic, dielectric, and mass transport behaviors \cite{sarkar2019high,aamlid2023understanding,sur2024high,tanveer2024synthesis}. Consequently, these materials serve as viable candidates for multifunctional applications across aerospace \cite{Dada19,DADA202143,ZHUO20241097,MENON2025130091,Dixit2022,xie2021role,xie2022phase}, biomedical applications\cite{RASHIDYAHMADY2023,Liu2022-lo,met12111940,Liu2022-qa,Junyi,RASHIDYAHMADY2023100009}, and energy storage applications \cite{FAN2025103954,Ouyang2024,D3YA00319A}.

Anomalous charge-transfer behavior in high-entropy materials (HEMs) has been widely reported in the literature. Representative examples include charge redistribution among constituent elements in high-entropy alloy (HEA) nanoparticles \cite{xu2020elemental}, the unconventional charge-transfer characteristics of Cu ions in the high-entropy oxide $\mathrm{(Mg_{0.2}Co_{0.2}Ni_{0.2}Cu_{0.2}Zn_{0.2}O)}$ \cite{sun2021high}, and anomalous hydrogen occupation of octahedral sites in FCC HEAs arising from electron localization between hydrogen and transition-metal atoms \cite{hu2022origin}.
In concentrated solid solutions, atoms frequently exhibit anomalous charge states that cannot be explained solely on the basis of isolated-atom electronegativity \cite{nohring2019correlation, korzhavyi2009electronic}. In certain cases, nominally electronegative elements behave as electron donors, whereas electropositive species accept charge. Such observations underscore the strongly environment-dependent nature of the electronic structure in real alloys and reveal the limitations of conventional charge-transfer heuristics. Recent density functional theory studies have further shown that charge redistribution in high-entropy alloys is governed by a complex interplay among local elastic strain, fluctuating Madelung-like electrostatic fields, and site-potential variations \cite{karabin2022ab,ho2023mechanism}, as well as electronic band filling effects \cite{zhao2019local}. Furthermore, defects in crystalline solids give rise to charge-density perturbations that may extend over several unit cells \cite{ghosh2022spectral}.

It is worth noting that the electronegativity can vary with the chemical and mechanical pressures \cite{rahm2019squeezing, dong2022electronegativity, sessa2022electronegativity}. Given its increasing use as a descriptor in data-driven materials design \cite{witman2023towards,zerdoumi2024combinatorial,zhao2024space,woods2023method,zhang2020dramatically,oshiya2024role,cao2025electronic}, a detailed understanding of the influence of chemical and mechanical pressure on charge-transfer behavior is essential.
Importantly, most of these studies of anomalous charge transfer focus on chemically homogeneous, defect-free crystals \cite{oh2021element,odbadrakh2019electronic,kohyama2021ab}, whereas real structural alloys are intrinsically defect-rich. 

Among crystalline defects, edge dislocations are particularly significant due to their long-range elastic fields and strong local perturbation of atomic coordination. An edge dislocation introduces a heterogeneous strain field characterized by compressive stress above the slip plane and tensile stress below it, leading to asymmetric modification of interatomic distances, local volume, and coordination environment. These effects directly influence the local electronic structure and, consequently, the direction and magnitude of charge transfer \cite{li2002effects}. Furthermore, the strain field associated with an edge dislocation alters local band structure through a combination of chemical pressure and coordination effects. Atoms in the compressive region experience increased orbital overlap and upward shifts in electronic energy levels, while atoms in the tensile region exhibit reduced overlap and lowered electronic states \cite{li2002effects}. This spatially varying electronic landscape breaks the assumption of uniform effective electronegativity within an alloy and gives rise to dislocation-induced charge polarization. As a result, charge transfer near an edge dislocation core can be qualitatively different from that in the bulk, with the same atomic species acting as a charge donor in one region and an acceptor in another. Dislocation-mediated anomalous charge transfer has important implications for solute segregation \cite{guo2025segregation} and dislocation mobility \cite{esfandiarpour2022edge,li2023fluctuations,sboui2025pins,liang2025computational}, and has been associated with phenomena such as solid-solution strengthening \cite{varvenne2016theory,varvenne2017solute,li2025universal,sboui2025pins}, hydrogen trapping \cite{marques2021effect}, and corrosion initiation \cite{wang2022influence}.
Importantly, edge dislocations provide a natural framework for probing non-classical charge-transfer behavior because they generate a continuous gradient of local atomic environments within a single crystal. The strain field is highly concentrated near the dislocation core and gradually decays toward the equilibrium lattice state with increasing distance from the core. Unlike grain boundaries or interfaces, dislocations enable a systematic correlation between charge transfer and well-defined elastic stress fields, thereby providing a mechanistic understanding of the coupling between strain and electronic structure in alloys. From this perspective, anomalous charge transfer should not be regarded as an isolated phenomenon, but rather as an intrinsic feature of defect-mediated alloy behavior.

In this work, we investigate anomalous charge transfer in Face Centered Cubic High Entropy Alloys in the presence of edge dislocations using first-principles electronic structure calculations. By explicitly resolving the spatial variation of atomic charge states across the dislocation strain field, we demonstrate how elastic distortion modifies effective electronegativity and drives non-intuitive charge redistribution. Our results establish a unified framework linking dislocation mechanics, electronic structure, and charge transfer, providing new insights into defect-controlled properties of complex alloys. The understanding of edge dislocation mediated charge transfer can provide the rigorous understanding of the evolution of the atomic scale distortion, edge dislocation mobility and catalytic behavior of complex alloys. 
The remainder of this paper is arranged as follows. In Section \ref{Sec:Methods} we present a detailed discussion on the methods used for our work, where we first discuss the generation of disordered supercells with and without dislocations, followed by details of electronic structure calculations, and finally present the definitions of charge and bond disproportions. In Section \ref{Sec:results} we first present the effect of bond disproportion on charge transfer followed by a discussion on the effect of the stress field charge transfer. Finally in Section \ref{Sec:Conclusion} we provide concluding remarks and outlook.

\section{Methods}\label{Sec:Methods}\label{Sec:Methods}
\subsection{Generation of supercell with disorder and dislocation}
We generated chemically disordered alloy supercells containing 1620 atoms using the disordered supercell generation functionality implemented in the OPERA framework \cite{anand2023order}. The OPERA (Order Parameter Engineering for RAndom Systems) approach constructs chemically disordered configurations by minimizing an order parameter $\Lambda$, defined as,
\begin{equation}
{
    \Lambda = \sum_{i \neq j}^{K_1}\left(1 -  \frac{m_{ij}}{2n} \right)+ \sum_{i \neq j}^{K_2}\left(\frac{m_{ij}}{n} - 1\right)   
    }
\end{equation}
where, ${K_1}$ and ${K_2}$ are the number of type of unlike and like bonds, respectively. Here ${m_{ij}}$ and ${m_{ii}}$ represent the numbers of unlike and like bonds in the generated supercell, while ${n}$ corresponds to the number of like bonds in a perfectly disordered configuration and is given by ${\frac{N}{2K_1 + K_2}}$ with ${N}$ being the number of atoms in the supercell. The OPERA combinatorial sampling procedure generates supercells satisfying ${\Lambda = 0}$.   
Five independent configurations were generated for each alloy composition. An edge dislocation dipole was subsequently introduced into each configuration, followed by energy minimization using published classical interatomic potentials for CoNi \cite{beland2016features}, CoCrNi \cite{li2019strengthening}, and CoCrFeMnNi \cite{choi2018understanding} within LAMMPS \cite{thompson2022lammps}. The edge dislocation dipole was inserted into the FCC supercells using Atomsk \cite{hirel2015atomsk}. The dipole configuration ensures a zero net Burgers vector under periodic boundary conditions, thereby eliminating long-range elastic fields, yielding finite elastic energy, and stabilizing the system for reliable structural relaxation and accurate characterization of dislocation core properties \cite{rodney2017ab}. Following the introduction of the dislocation dipole, the supercells contained 1684 atoms. After energy minimization, the atomic pressure at each atomic site was obtained from the local stress tensor computed in LAMMPS.

\subsection{Electronic structure calculations}
Large-scale \emph{ab initio} calculations were performed for chemically disordered Face Centered Cubic (FCC) CoNi, CoCrNi, and CoCrFeMnNi alloys using the locally self-consistent multiple scattering (LSMS) method \cite{wang1995order,eisenbach2017gpu}. LSMS is a real-space, Green’s function based electronic-structure approach derived from the Korringa–Kohn–Rostoker (KKR) multiple-scattering formalism \cite{wang1995order}. In conventional KKR theory, the electronic structure is obtained through inversion of a global scattering matrix, leading to a computational cost that scales cubically with system size, $O(N^{3})$, which becomes prohibitive for chemically complex large-scale systems.

LSMS circumvents this limitation through a local self-consistency framework. For each atomic site, a finite local interaction zone (LIZ), containing the surrounding near-neighbor environment, is constructed independently. The multiple-scattering problem is solved within each LIZ to obtain the local Green’s function, from which the site-resolved charge density, magnetic moments, and effective potentials are determined. Self-consistency is achieved locally at every atomic site, while a global Fermi level is maintained to ensure overall charge conservation. Because the size of the LIZ remains fixed, the computational effort per atom is effectively constant, resulting in linear $O(N)$ scaling with system size. This formulation is naturally parallelizable and is therefore well suited for first-principles simulations of large chemically disordered supercells in high-entropy alloys.

\begin{figure}
    \centering
    \includegraphics[width=\linewidth]{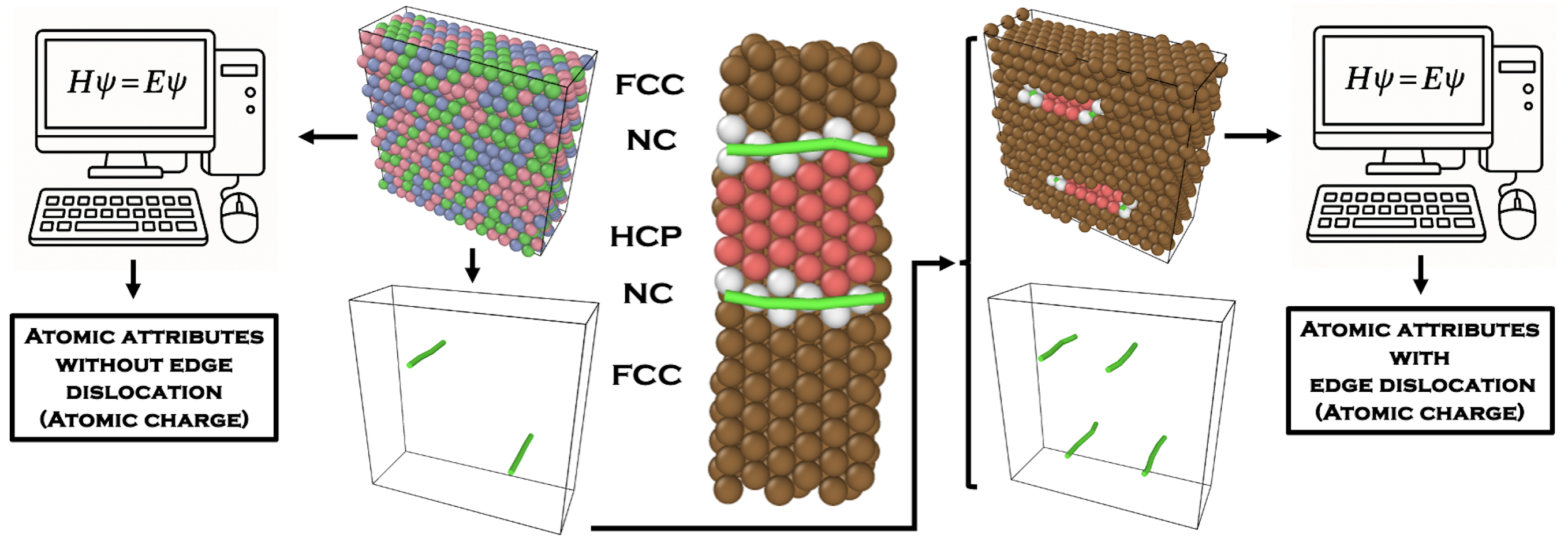}
    \caption{Schematic representation of the method for studying the influence of the edge dislocation on charge transfer characteristics in alloys, including CoNi, CoCrNi, and CoCrFeMnNi.}
    \label{fig:method}
\end{figure}

The electronically minimized structures, both with and without edge dislocations, were subsequently analyzed to quantify the influence of dislocation strain fields on local atomic charge transfer in the alloys (Fig. \ref{fig:method}).

\subsection{Atomic charge disproportion and bond disproportion vector}
To quantify local deviations from ideal random chemical mixing in multicomponent alloys, we employ the bond disproportion vector (${\bm{\lambda}}$), which characterizes variations in nearest-neighbor bond statistics relative to a perfectly random reference state. For a given bond type ${i}$, let ${B_i^{(1\mathrm{NN})}}$ denote the number (or probability) of bonds in the first-nearest-neighbor shell, and ${B_i^{(\mathrm{rand})}}$ represent the corresponding expectation for an ideal random alloy of identical composition. The bond disproportion associated with bond type ${i}$ is then defined as

\begin{equation}
\lambda_i = B_i^{(1\mathrm{NN})} - B_i^{(\mathrm{rand})}.
\end{equation}

The complete bond disproportion vector is expressed as

\begin{equation}
{\bm{\lambda}} =
\left(
\lambda_1, \lambda_2, \ldots, \lambda_N
\right),
\end{equation}

where ${N}$ denotes the total number of bond types considered in the supercell. Positive values of ${\lambda_i}$ indicate an enhancement of the corresponding bond population relative to the random alloy limit, whereas negative values signify bond depletion. The vector ${\bm{\lambda}}$ therefore provides a compact statistical descriptor of local chemical-environment fluctuations in chemically disordered alloys.

To quantify the influence of local chemical environments on charge redistribution, we define the charge disproportion parameter, ${\Delta_q}$, as the difference between the atomic charge in the alloy, ${Z_{\mathrm{alloy}}}$, and the charge of the corresponding elemental atom, ${Z_{\mathrm{atom}}}$,

\begin{equation}
\Delta_q =
Z_{\mathrm{alloy}} - Z_{\mathrm{atom}}.
\end{equation}

A positive value of ${\Delta_q}$ corresponds to net charge gain, whereas ${\Delta_q < 0}$ indicates charge depletion. The dependence of ${\Delta_q}$ on the bond disproportion vector ${\bm{\lambda}}$ provides a quantitative measure of how variations in the local coordination environment influence charge transfer at a given atomic site.

\section{Results and discussion}\label{Sec:results}
\subsection{{Effect of bond disproportion on charge transfer}}
\begin{figure}
    \centering
    \begin{subfigure}[b]{0.4\textwidth}
        \centering
        \includegraphics[width=0.75\linewidth]{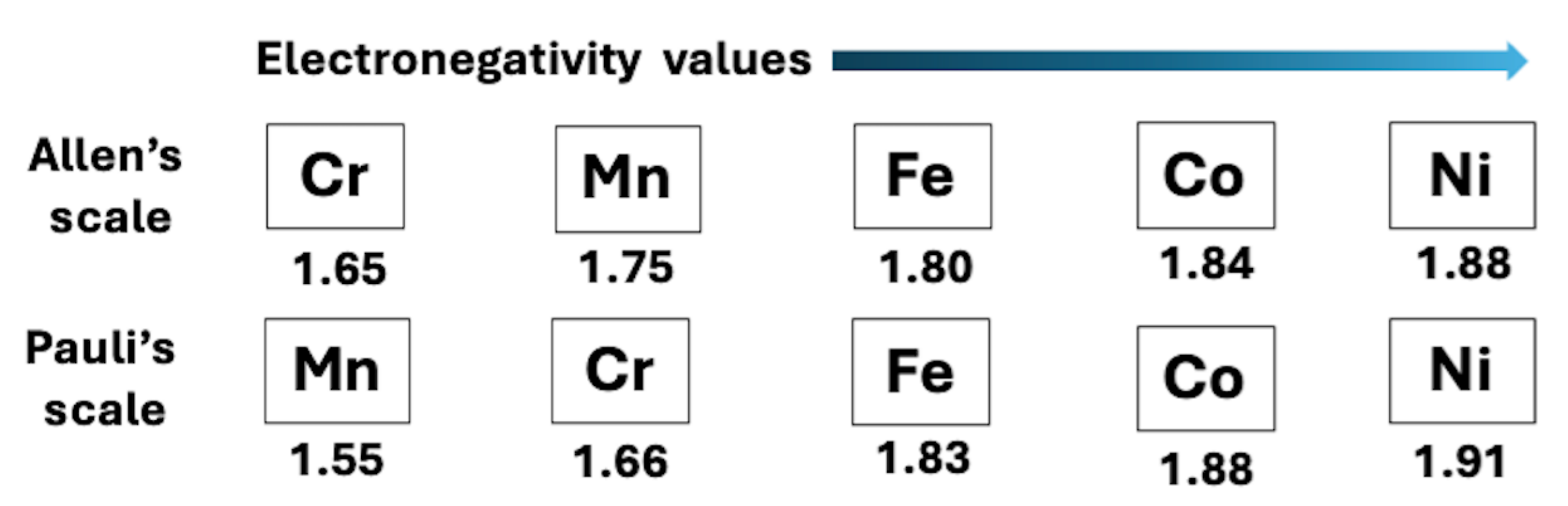}
        \includegraphics[width=\linewidth]{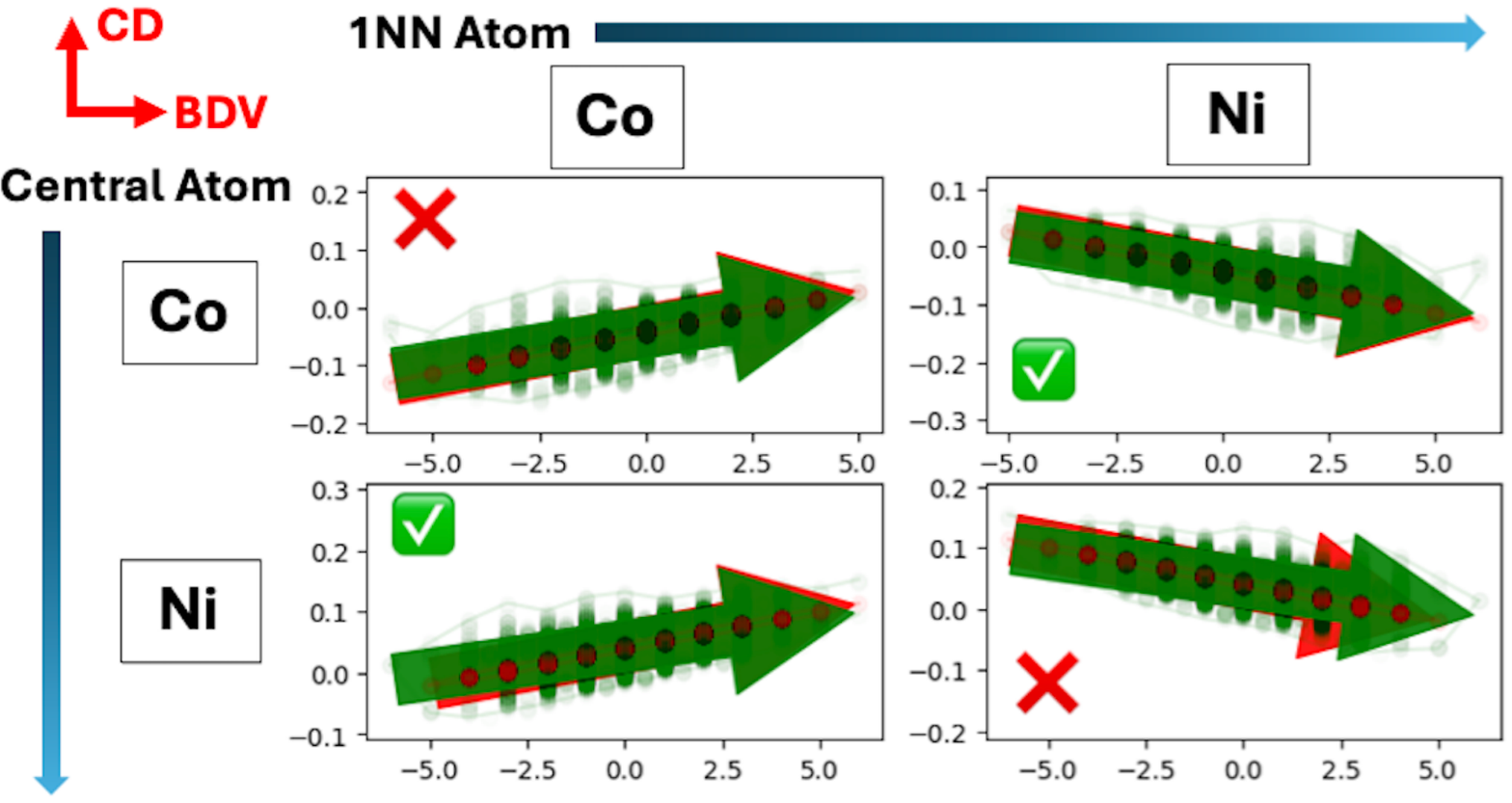}
        \caption{CoNi}
        \label{fig:bdvVsCD-CoNi}
    \end{subfigure}
    \begin{subfigure}[b]{0.59\textwidth}
        \includegraphics[width=\linewidth]{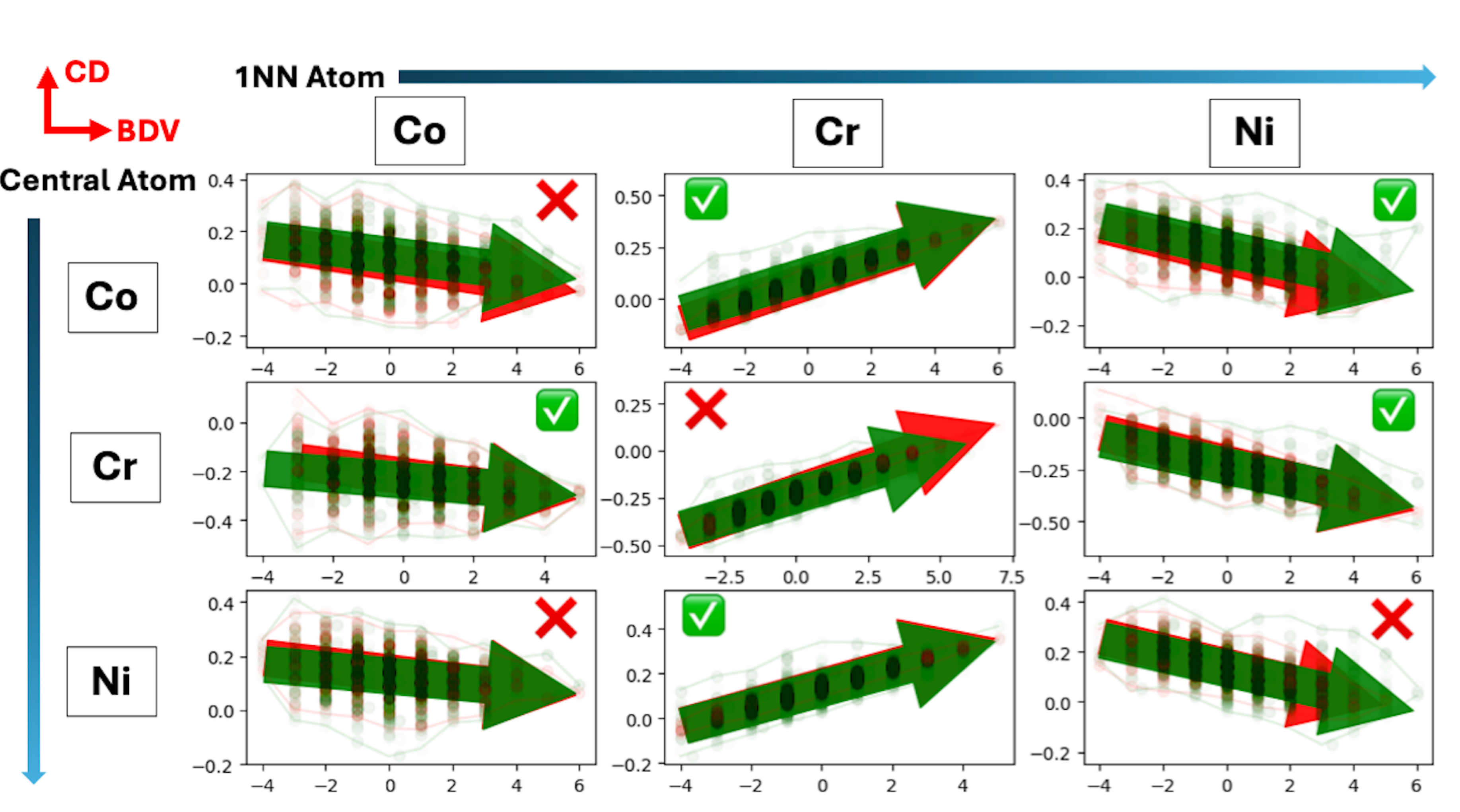}
        \caption{CoCrNi}
        \label{fig:bdvVsCD-CoCrNi}
    \end{subfigure}
    \begin{subfigure}[b]{\textwidth}
        \includegraphics[width=\linewidth]{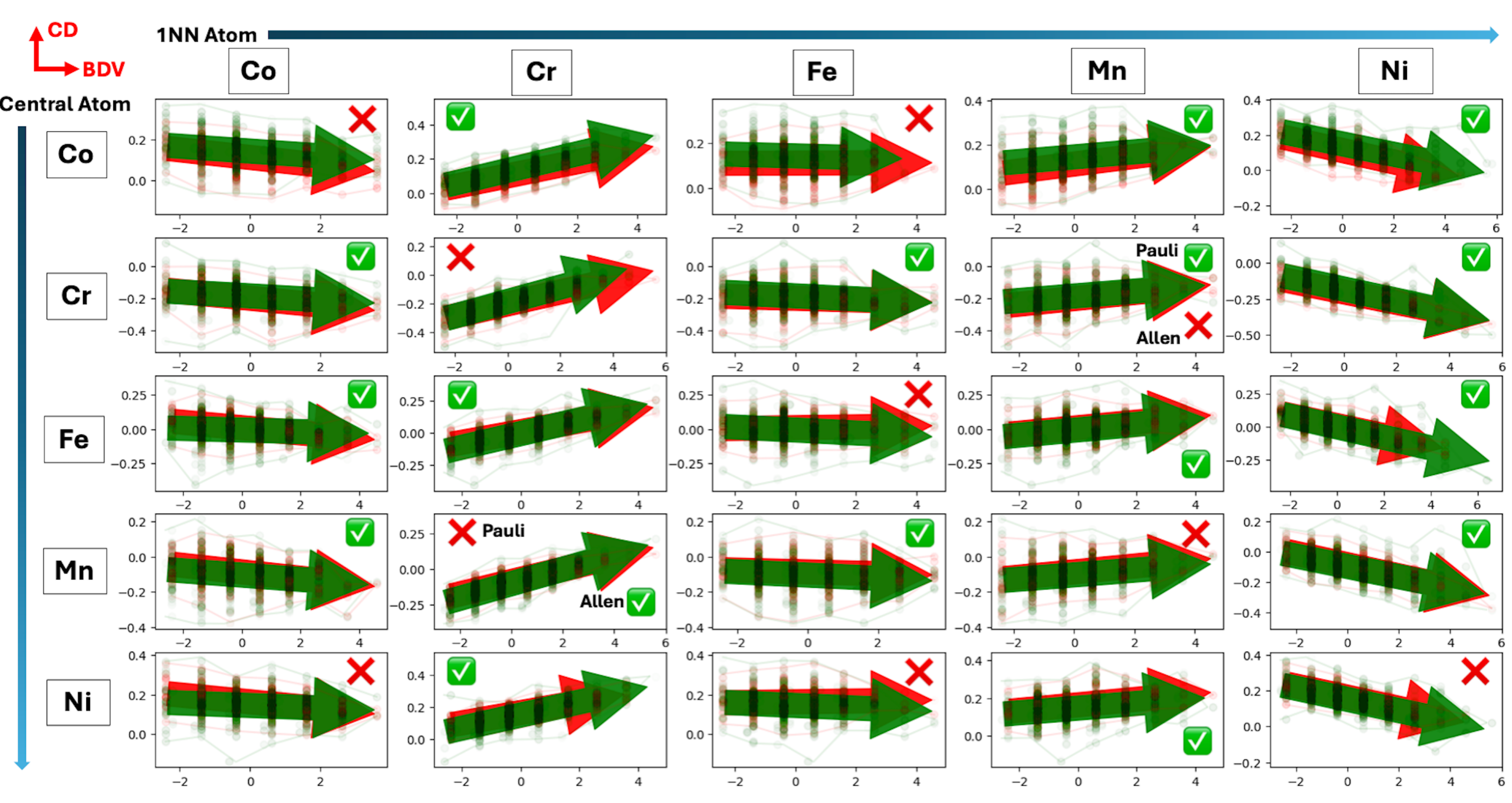}
        \caption{CoCrFeMnNi}
        \label{fig:bdvVsCD-CCFMN}
    \end{subfigure}

    \caption{Top panel shows the electronegativity values of the individual atomic species. Bottom panel shows the the variation of the charge disproportion ($\Delta_q$) with bond disproportion vector ($\bm{\lambda}$) entry for an atom in the center of the first nearest-neighbor (1NN) coordination shell for (a) CoNi, (b) CoCrNi and (c) CoCrFeMnNi. The green and red arrow shows the linearly fitted trend of $\Delta_q$ with respect to $\bm{\lambda}$ without and with edge dislocation in the supercell, respectively. It can be seen that the qualitative trend of the bond disproportion on the central atom is dependent only on the bond disproportion in first nearest neighbor atomic species (1NN) and not on the identity of the central atom}.
    \label{fig:arrow-cd-bdv}
\end{figure}

Figure \ref{fig:bdvVsCD-CoNi} quantifies the dependence of charge disproportion on the local bond disproportion of the nearest-neighbor (1NN) coordination shell surrounding a reference (central) atom in the Co–Ni alloy. For a Co central atom, increasing the Ni number density in 1NN produces a monotonic decrease in charge disproportion on the Co central atom, signifying progressive electron depletion from Co. This behavior reflects enhanced electronic screening by the more electronegative Ni atoms, which preferentially draw charge from the central Co through altered local bonding environments. When the central atom is Ni and Co is varied in 1NN, an opposite trend is observed.

The electronegativity hierarchy of the constituent elements, shown in Fig. \ref{fig:arrow-cd-bdv}, provides a first-order rationalization of these trends, as the species with higher electronegativity acts as a charge acceptor. However, the observed response cannot be interpreted solely on the basis of pairwise electronegativity differences. Modifying the NN1 composition inherently redistributes the local chemical environment, such that an increase in the number density of one species is accompanied by a compensating decrease in the other.

A key mechanistic insight from Fig. \ref{fig:bdvVsCD-CoNi} is that bond disproportion associated with a given atomic species induces qualitatively opposite charge responses depending on whether the central atom is chemically identical to, or distinct from, the species driving the bond disproportion. Although this behavior appears counterintuitive when considering isolated-atom electronegativities, it emerges naturally once global charge conservation within the supercell is enforced. Excess Ni in the NN1 shell drives charge accumulation consistent with its higher electronegativity, thereby increasing electron depletion from a central Co atom. Simultaneously, this excess necessarily corresponds to a deficit of Co neighbors, yielding negative bond disproportion values. Therefore, the resulting charge loss cannot be attributed to electronegativity arguments alone, but instead arises from collective electronic redistribution constrained by supercell-level charge neutrality.

Next, we take the case of the medium-entropy CoCrNi alloy. Figure \ref{fig:bdvVsCD-CoCrNi} shows the effect of bond disproportion on the charge disproportion for this alloy. As the number of atoms in the alloy increases, the deviation from the electronegativity trend becomes more pronounced.  Apart from the deviation from the electronegativity trend for a similar central atom and 1NN case as discussed previously, the deviation for charge disproportion on Ni central atom with bond disproportion of Co atom is evident.

Finally, we consider the CoCrFeMnNi high-entropy alloy. Figure \ref{fig:bdvVsCD-CCFMN} shows that the primary deviations occur for combinations involving similar central atoms and first-nearest-neighbor (1NN) bond pairs. Additional deviations from the electronegativity trend are observed for Ni-centered environments with Co 1NN bond descriptors, Co-centered environments with Fe 1NN bond descriptors, and Ni-centered environments with Fe 1NN bond descriptors. Deviations are also evident for Mn-centered environments with Cr 1NN bond descriptors when the Pauli electronegativity scale is used, and for Cr-centered environments with Mn 1NN bond descriptors when the Allen electronegativity scale is considered.

Several observations emerge from the analysis of the binary CoNi, ternary CoCrNi, and quinary CoCrFeMnNi high-entropy alloy systems. First, the identity of the central atom does not significantly affect either the qualitative or statistical variation of the charge disproportion on the central atom as a function of the bond disproportion associated with different first-nearest-neighbor (1NN) species. Moreover, the charge disproportion induced by specific 1NN atomic species exhibits similar qualitative trends across different local environments, which naturally gives rise to deviations from conventional electronegativity-based expectations.

Electronegativity is fundamentally defined within a pairwise elemental framework, wherein electronic charge is transferred toward the more electronegative species. However, in extended solid solutions such as concentrated alloys, this pairwise description appears insufficient, as demonstrated in Fig. \ref{fig:arrow-cd-bdv}. In these systems, the simultaneous requirements of charge neutrality and overall charge conservation govern the charge equilibration process, thereby leading to the observed anomalous charge-transfer behavior.

The deviations from conventional electronegativity trends can be interpreted at the atomistic scale, particularly through the relationship between the charge disproportion ($\Delta_q$) of an atom and the corresponding bond disproportion (BD). Nevertheless, from a statistical perspective, electronegativity trends still provide an appropriate qualitative description of the overall charge disproportion behavior. For example, in the quinary CoCrFeMnNi high-entropy alloy, Co and Ni predominantly exhibit electropositive character, whereas Cr and Mn display electronegative behavior. Fe remains statistically near neutral. In the ternary CoCrNi alloy, Co and Ni exhibit electronegative tendencies, while Cr behaves electropositively. In the binary CoNi alloy, Co and Ni display electropositive and electronegative behavior, respectively.

These observations suggest that, although the electronegativity concept remains qualitatively valid for describing average charge-transfer behavior in disordered concentrated alloys, it is insufficient for capturing the atomistic-scale charge disproportion arising from the complex local chemical environments present in such systems.

We note that Fig. \ref{fig:arrow-cd-bdv} presents the relationship between the charge disproportion ($\Delta_q$) of the central atom and the bond disproportion (BD) within the first-nearest-neighbor (1NN) shell, both in the presence and absence of an edge dislocation dipole. The introduction of the dislocation increases the scatter in the observed trends. This enhanced scatter in $\Delta_q$ can be attributed to the influence of the stress field associated with the edge dislocation.

\begin{figure}
    \begin{subfigure}[b]{\textwidth}
        \centering
        \includegraphics[width=0.8\linewidth]{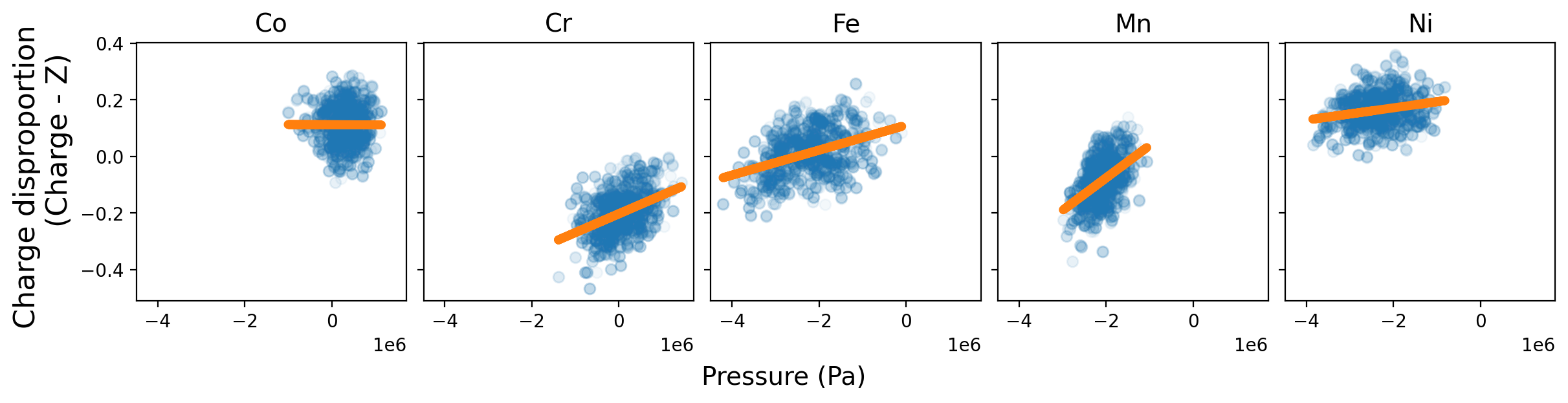}
        \includegraphics[width=0.8\linewidth]{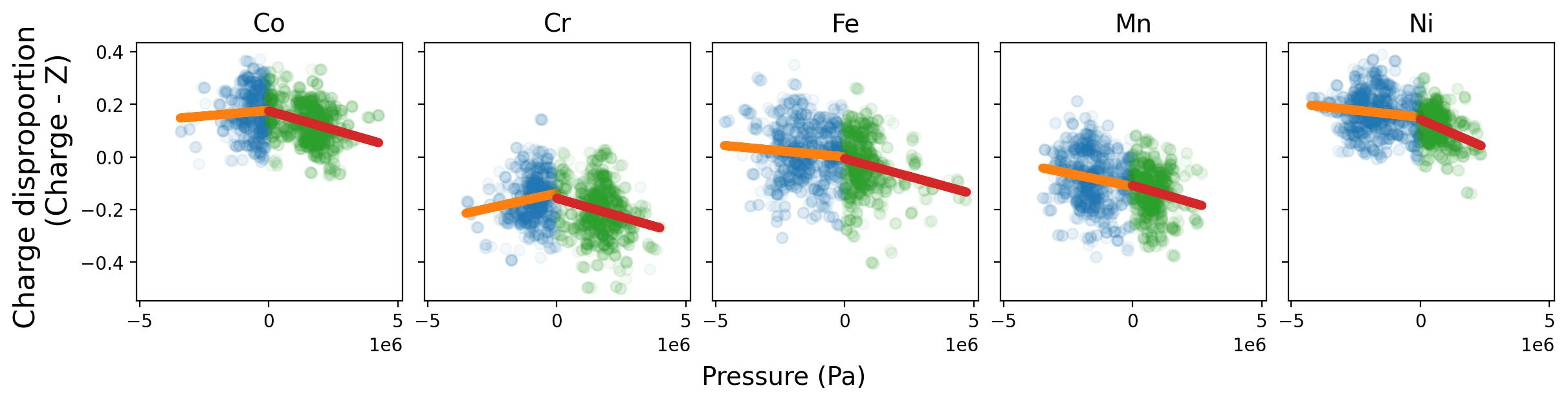}
        \includegraphics[width=0.8\linewidth]{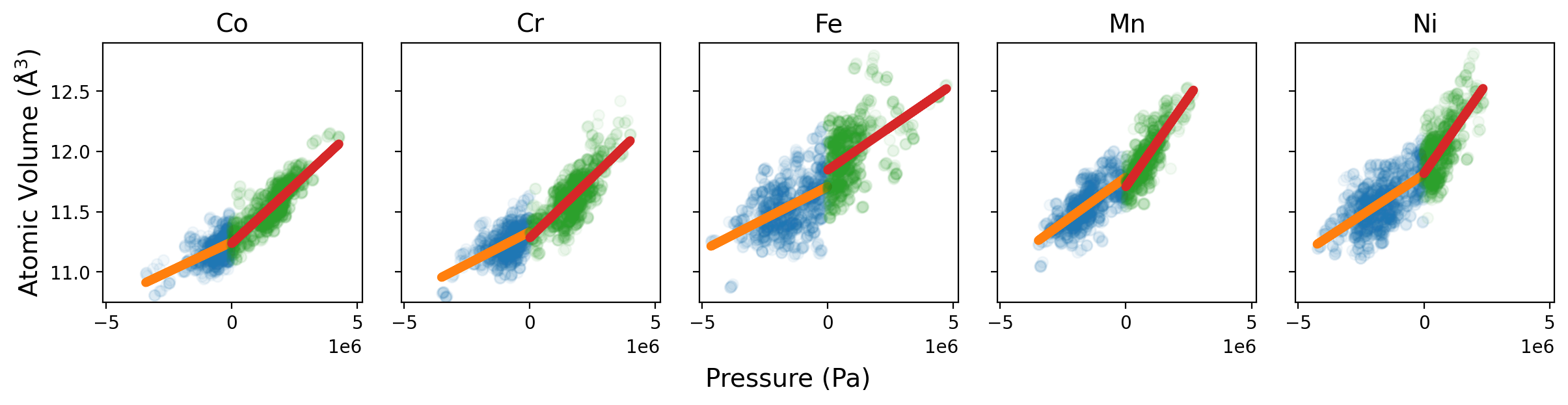}
        \caption{CoCrFeMnNi}
        \label{fig:chargedis_pres-CCFMN}
    \end{subfigure}
    \hfill
    \begin{subfigure}[b]{0.6\textwidth}
        \centering
        \includegraphics[width=0.8\linewidth]{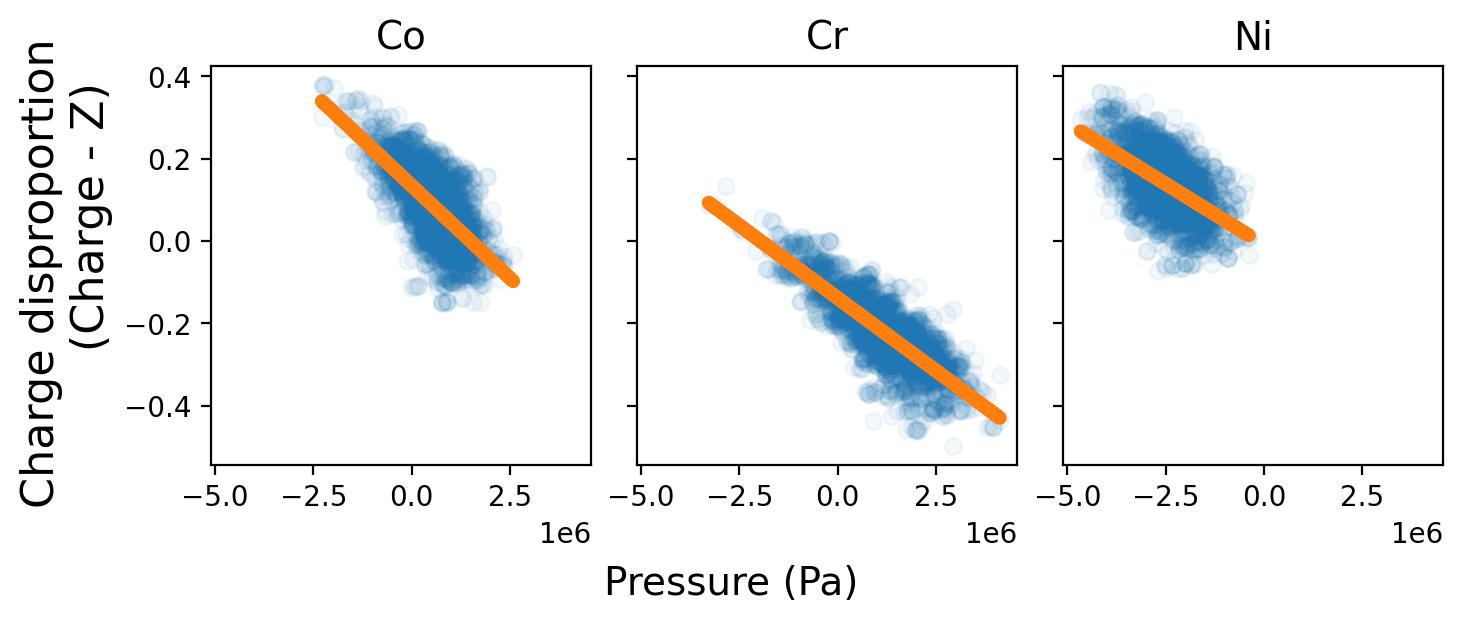}
        \includegraphics[width=0.8\linewidth]{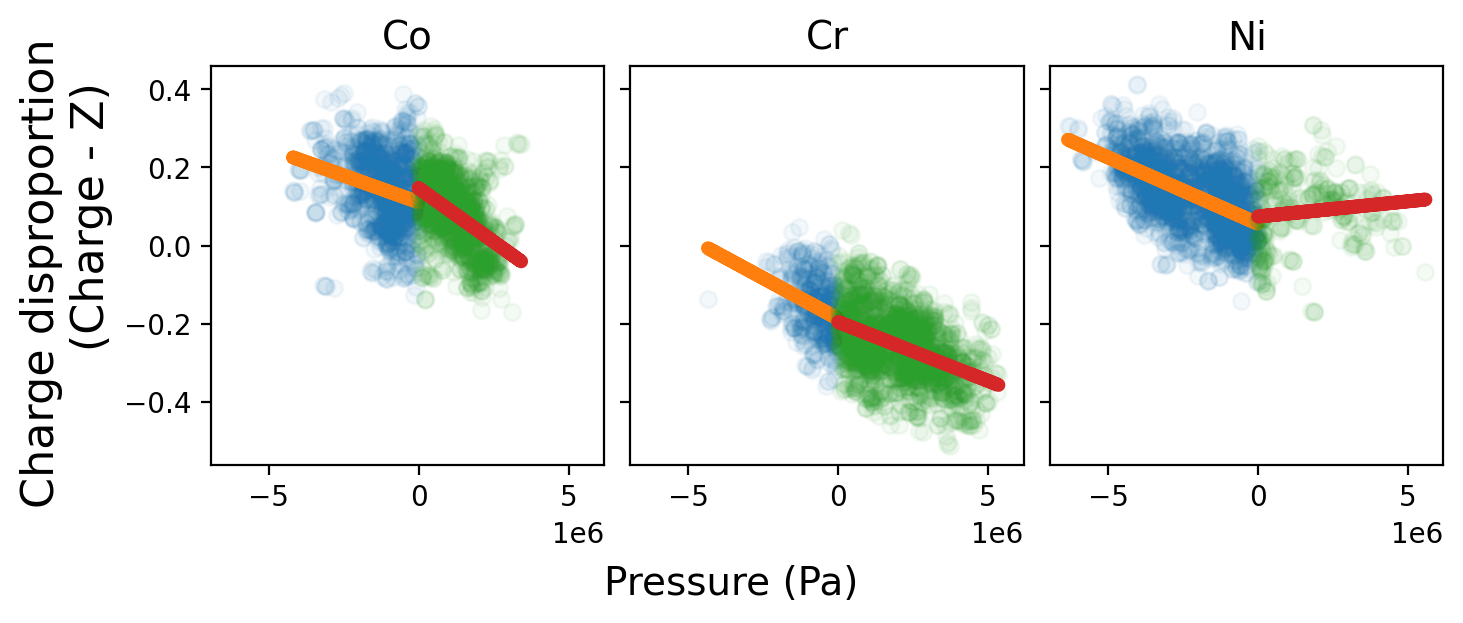}
        \includegraphics[width=0.8\linewidth]{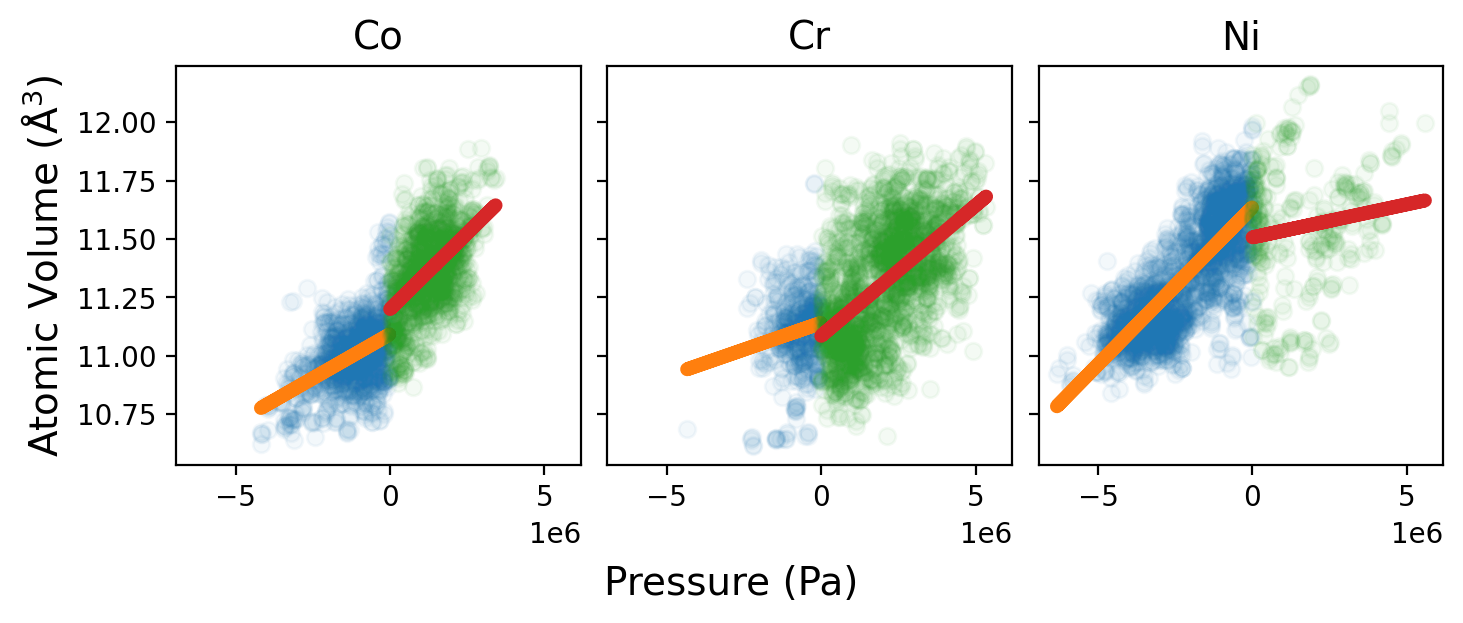}
        \caption{CoCrNi}
        \label{fig:chargedis-CCN}
    \end{subfigure}
    \hfill
    \begin{subfigure}[b]{0.4\textwidth}
        \centering
        \includegraphics[width=0.8\linewidth]{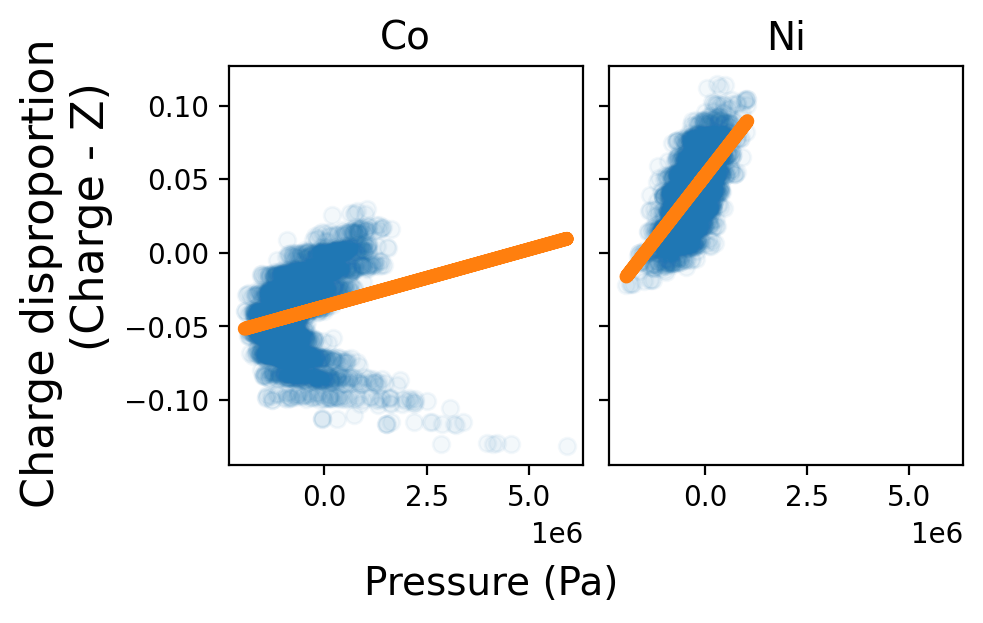}
        \includegraphics[width=0.8\linewidth]{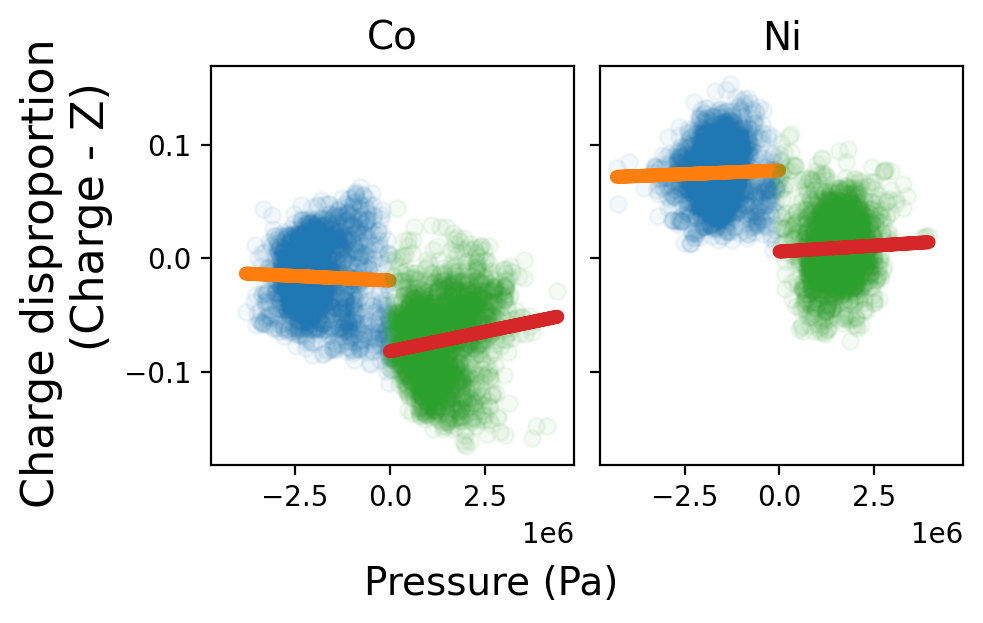}
        \includegraphics[width=0.8\linewidth]{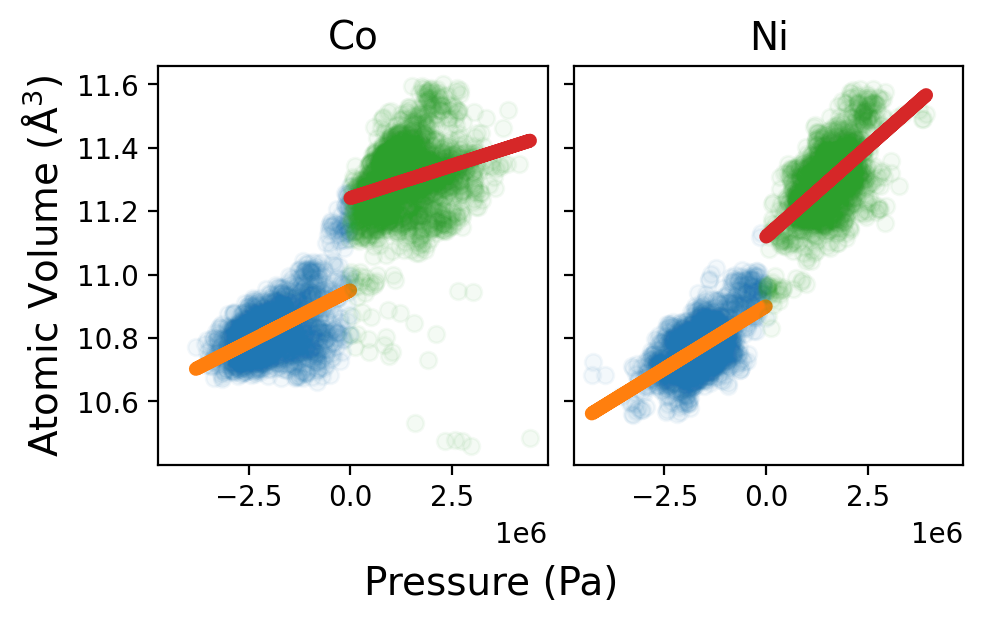}
        \caption{CoNi}
        \label{fig:chargedis-CoNi}
    \end{subfigure}
    
    \caption{Charge disproportion versus pressure for alloys. The top, middle and bottom panel shows the charge disproportion for elements in alloys without dislocation,  with dislocation and atomic volume with dislocation, respectively. The lines on the data has been drawn for the visual aid to approximate the different trends as a function of sign of the pressure.}
    \label{fig:all-volume}
\end{figure}

Figure \ref{fig:all-volume} further illustrates the variation of charge disproportion, with and without dislocation, together with the corresponding atomic volumes in the presence of dislocations for the CoCrFeMnNi, CoCrNi, and CoNi alloys. In the case of the CoCrFeMnNi high-entropy alloy (Fig. \ref{fig:chargedis_pres-CCFMN}), the charge disproportion of the constituent elements exhibits a positive dependence on pressure. The introduction of the edge dislocation dipole generates both tensile and compressive stress regions, resulting in distinct slopes in the charge disproportion--pressure relationship. Correspondingly, the atomic volumes also display different trends in the tensile and compressive regions.

For the CoCrNi alloy (Fig. \ref{fig:chargedis-CCN}), the variation of $\Delta_q$ in the defect-free system differs substantially from that observed in CoCrFeMnNi. The presence of the edge dislocation dipole significantly modifies the $\Delta_q$ behavior of Ni under both compressive and tensile stresses. A similar change in slope with stress state is also observed in the atomic volume of Ni in the presence of the dislocation dipole. In contrast, the CoNi alloy exhibits no substantial difference between Co and Ni in either the charge disproportion behavior, with and without dislocation, or the variation of atomic volume in the presence of the edge dislocation dipole.

\begin{figure}
    \begin{subfigure}[b]{\textwidth}
        \centering
        \includegraphics[width=\linewidth]{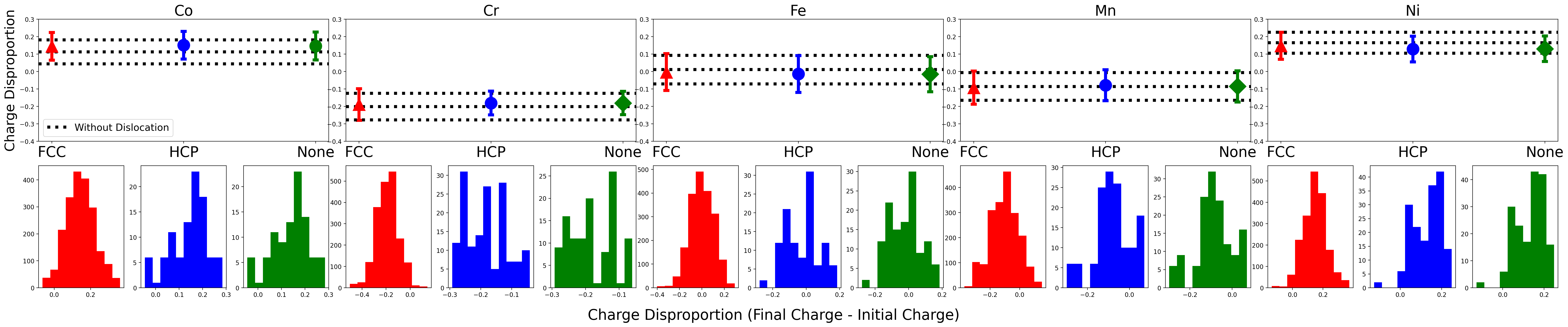}
        \includegraphics[width=\linewidth]{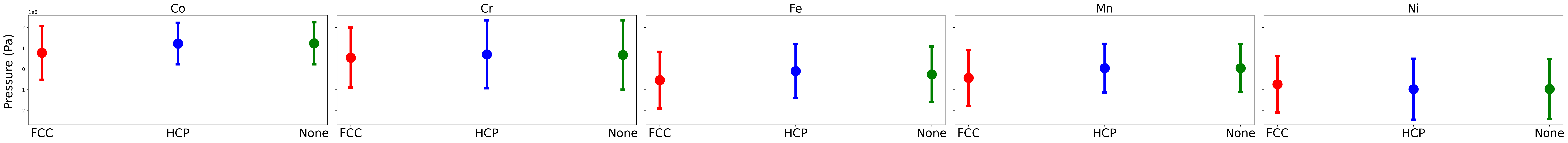}
        \caption{CoCrFeMnNi}
        \label{fig:charge-CCFMN}
    \end{subfigure}
    \hfill
    \begin{subfigure}[b]{0.6\textwidth}
        \centering
        \includegraphics[width=\linewidth]{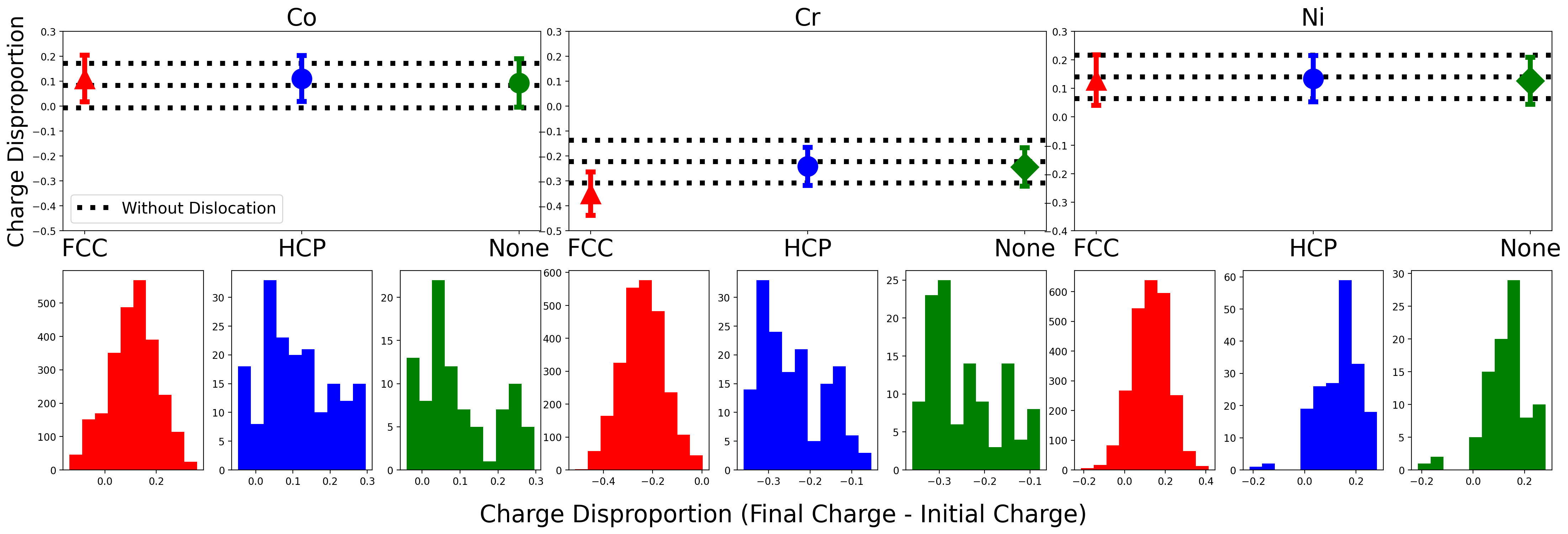}
        \includegraphics[width=\linewidth]{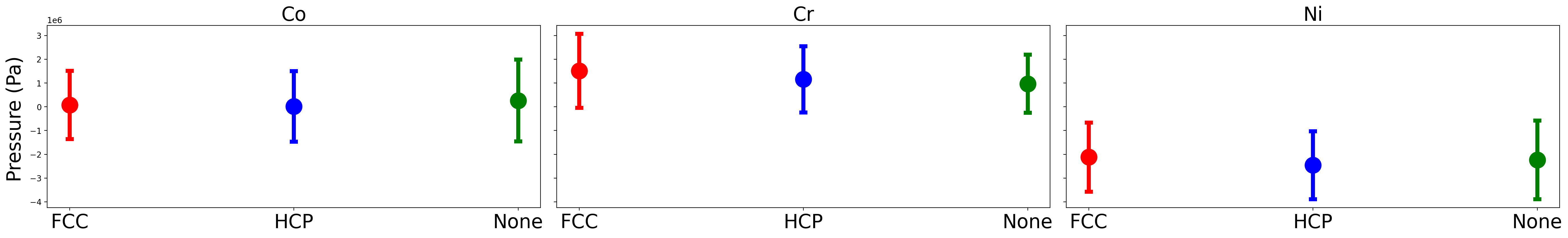}
        \caption{CoCrNi}
        \label{fig:charge-CCN}
    \end{subfigure}
    \hfill
    \begin{subfigure}[b]{0.4\textwidth}
        \centering
        \includegraphics[width=\linewidth]{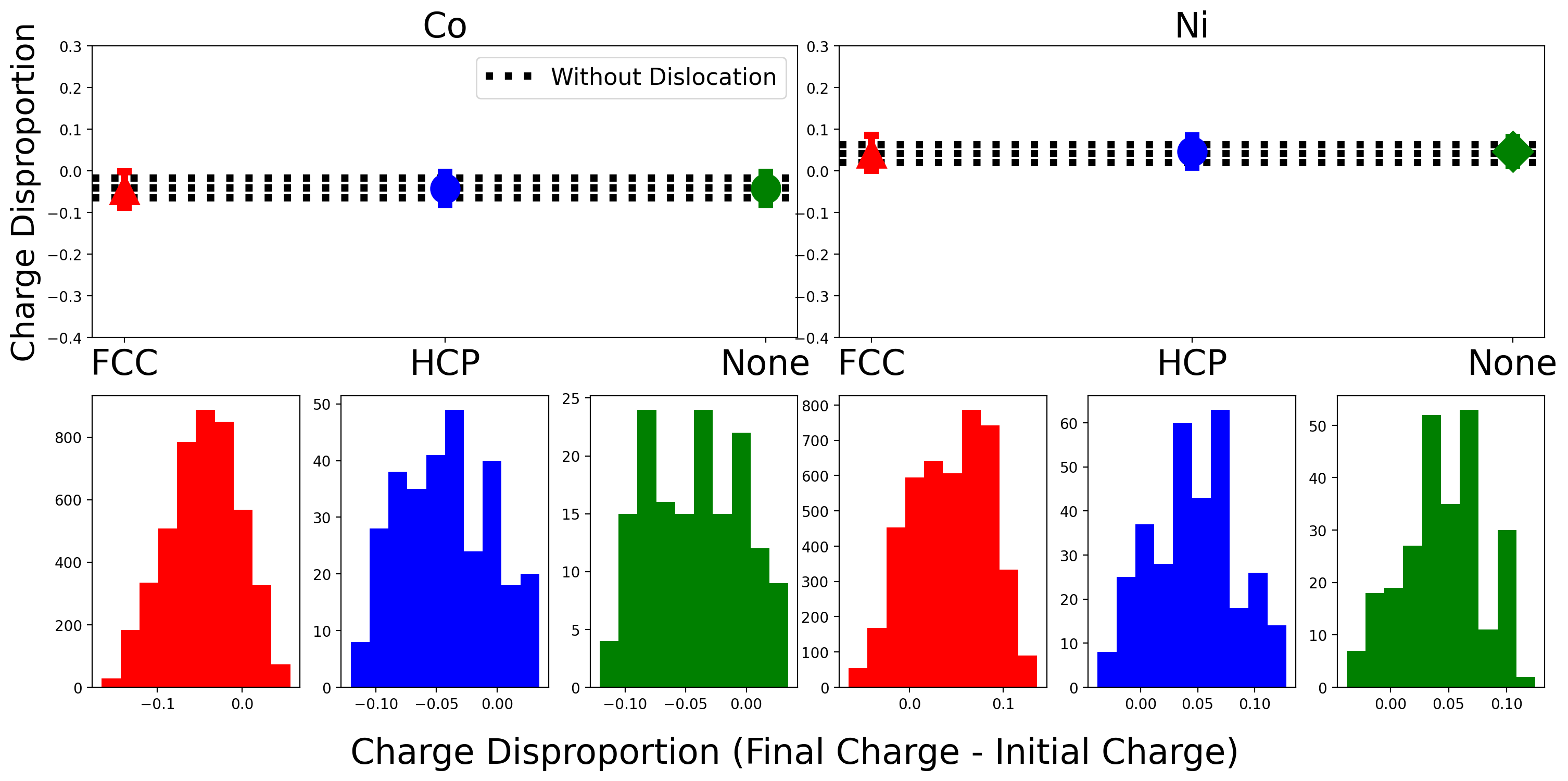}
        \includegraphics[width=\linewidth]{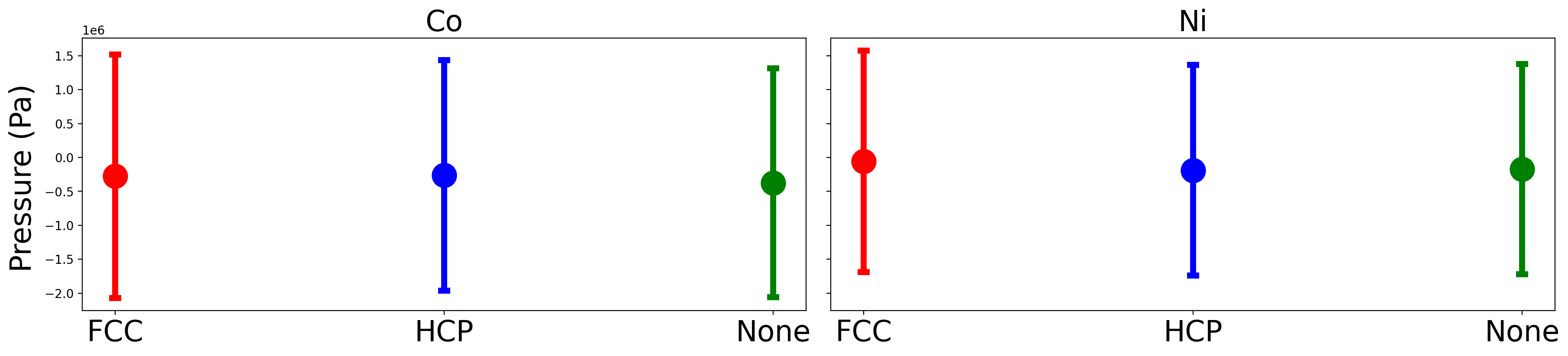}
        \caption{CoNi}
        \label{fig:charge-CoNi}
    \end{subfigure}
    
    \caption{The variation of charge disproportion on elements in CoCrFeMnNi, CoCrNi and CoNi. The top panel shows the variation of charge disproportion in the supercell with edge dislocation dipole compared with the supercell without a dislocation dipole. The middle line represents the average, and the top and bottom lines represent the standard deviation added and subtracted from the mean values, respectively. The middle panel shows the discrete histogram for FCC, HCP, and non-coordinated atoms, and the bottom panel shows the atomic pressure (\emph{trace of stress tensor}) for atoms in FCC, HCP, and non-coordinated.}
    \label{fig:all-charge}
\end{figure}

Figure \ref{fig:all-charge} presents the variation of charge disproportion ($\Delta_q$) for FCC atoms, HCP atoms located between the dislocation partials, and non coordinated (NC) atoms situated near the dislocation core for the different alloy systems. The scatter in charge disproportion is substantially larger in CoCrNi and CoCrFeMnNi high-entropy alloy than in the CoNi alloy.

Another important observation is that the FCC, HCP, and NC atomic populations statistically experience comparable ranges of stress across all alloy systems. Similarly, the range of charge disproportion associated with each elemental species remains statistically comparable across these local environments. This finding is significant because it indicates that the presence of the edge dislocation does not induce a distinctly different charge disproportion behavior for atoms located in different dislocation-related regions.

\subsection{{Effect of edge dislocation stress field on charge transfer}}

The presence of dislocations modifies the characteristics of charge transfer through its influence on the electron work function (EWF). Lattice distortions induced by the dislocation alter the depth of local electron potential wells, thereby changing the EWF \cite{li2002effects}. Therefore, the variation in EWF ($\Delta \mathrm{EWF}$) should be considered separately for tensile and compressive stress states, since an edge dislocation generates both.

Compressive strain reduces interatomic spacing, resulting in deeper potential wells, whereas tensile strain produces shallower wells. The latter facilitates enhanced charge mobility compared to the compressive case due to the reduced confinement of electronic states. Overall, the presence of dislocations leads to a reduction in the system-wide EWF relative to the defect-free lattice. This reduction provides a consistent explanation for the increased scatter and dispersion in charge disproportion observed in the dislocation-containing system, as shown in Fig. \ref{fig:arrow-cd-bdv}.

To rationalize the effect of dislocation splitting in FCC crystal structures on charge-transfer characteristics, we consider a crystalline system with point nuclei located at positions ${R_i}$. The electrostatic potential at an electron position ${r}$ can be expressed as

\begin{equation}
\phi(r) = \sum_i \frac{Ze}{4\pi\epsilon_0 \lvert R_i - r \rvert}
\end{equation}

where ${Z}$ denotes the screened nuclear charge. In the presence of a dislocation, the resulting atomic displacement field ${u(r)}$ induces a perturbation in the electrostatic potential at site ${r}$, denoted $\Delta \phi({r})$, which can be approximated as

\begin{equation}
\Delta\phi (r) \approx \sum_i \frac{Ze}{4\pi\epsilon_0} \frac{u(R_i)\cdot(R_i-r)}{\lvert R_i - r \rvert^3}
\end{equation}

This expression indicates that tensile displacements reduce the effective Coulomb interaction, thereby lowering the electron work function (EWF), consistent with the discussion above. To further relate this effect to dislocation parameters, we express the displacement field in terms of the Burgers vector $\bf{b}$ and the lattice parameter ${a}$, using the scaling relations ${u \propto {\bf{b}}}$ and ${R \sim a}$. This leads to the approximate proportionality
\begin{equation}
\Delta \phi \propto \frac{Ze}{4\pi \epsilon_0}\cdot \frac{{\bf{b}}}{a^2}
\end{equation}

\begin{figure}
    \centering
    \includegraphics[width=\linewidth]{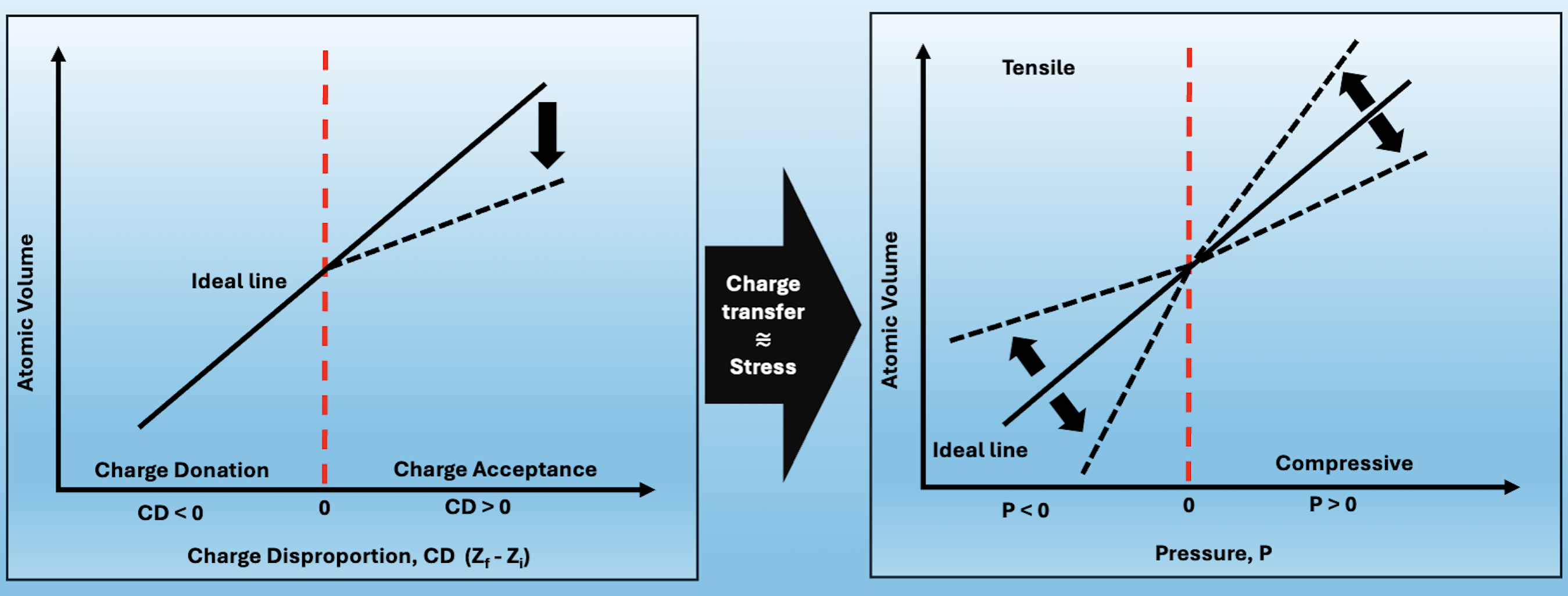}
    \caption{{A schematic representation of the variation in the expected atomic-volume with charge gain as well as charge loss (left panel) as represented with an ``ideal line". The deviation of the atomic-volume from the ideal line is also shown as a hatched line marked with arrow. The right panel shows the deviation in atomic volume with pressure.}}
    \label{fig:schematic}
\end{figure}

In the case of a split edge dislocation in FCC alloys, the local atomic environments can be categorized as non-coordinated (NC; atoms near the partial dislocation core, Fig. \ref{fig:method}), HCP-coordinated atoms located between the partials, and FCC-coordinated atoms in the bulk-like regions. On the basis of the associated lattice distortion, one would anticipate a hierarchy in the electrostatic potential perturbation such that $\Delta\phi_{\mathrm{NC}} > \Delta\phi_{\mathrm{HCP}} > \Delta\phi_{\mathrm{FCC}}$. However, the charge disproportion $\Delta q$, as shown in Fig. \ref{fig:all-charge}, does not follow this ordering for CoNi, CoCrNi, and CoCrFeMnNi high-entropy alloy, thereby indicating an anomalous charge-transfer response in these alloys. This deviation suggests that $\Delta q$ is not governed solely by the electrostatic potential variation but rather scales as $\Delta q \propto \Delta \mathrm{EWF} \cdot S$, where $S$ denotes the chemical softness. A higher chemical softness corresponds to a more easily deformable electronic cloud and thus enhanced charge redistribution capability. Furthermore, $S$ is pressure-dependent and increases under compressive stress \cite{dong2022electronegativity}. Consequently, tensile stress reduces $S$ and suppresses charge transfer, whereas compressive stress has the opposite effect, enhancing electronic reorganization and charge redistribution. The resulting charge redistribution $\Delta q$ couples directly to the atomic volume response, such as a net gain of electronic charge ($\Delta E > 0$ at a given site) promotes lattice expansion and induces tensile chemical stress, whereas charge depletion leads to local contraction and compressive chemical stress. This constitutive response is schematically represented by the “ideal line” in the left panel of Fig. \ref{fig:schematic}.

However, density functional theory calculations have demonstrated that in FCC alloys, Cr atoms surrounded by Co, Fe, Mn, or Ni follow electronegativity-driven charge transfer trends but exhibit an anomalous structural response, where charge accumulation does not result in the expected volume expansion; instead, a volume collapse is observed \cite{anand2020electron}. This counterintuitive behavior has been attributed to magnetic ground-state effects and spin fluctuations. When this coupling between charge transfer and chemical stress is extended to defect-containing alloys, the stress field of an edge dislocation further modifies the charge redistribution, leading to a non-trivial change in the slope of $\Delta q$ as a function of local stress state (Fig. \ref{fig:all-volume}). As illustrated in the right panel of Fig. \ref{fig:schematic}, deviations of atomic volume from the expected stress dependence can thus be traced back to this anomalous, environment-dependent charge-transfer response in concentrated alloys.

\section{Concluding remarks and outlook}\label{Sec:Conclusion}

In summary, this work demonstrates that charge-transfer behavior in extended solid solutions deviates from the predictions of conventional pair-interaction-based electronegativity frameworks, which are insufficient to capture even the qualitative features of atomic-scale charge redistribution. While electronegativity equalization remains a useful descriptor at the statistical level, the atomistic charge transfer in concentrated alloys is governed by a more complex, environment-dependent response.

The introduction of edge dislocations leads to a pronounced increase in the scatter of charge disproportion, which can be rationalized by a reduction in the electron work function induced by the dislocation strain field. Furthermore, both charge disproportion and atomic volume exhibit distinct scaling behaviors under tensile and compressive stress states associated with the dislocation.

Finally, FCC, HCP (located between partial dislocations), and non-coordinated atoms near the dislocation core display statistically similar anomalous charge-disproportion behavior. This observation can be understood in terms of the comparable distribution of local atomic stresses across these coordination environments at the atomistic scale.

Future work will focus on developing a predictive multiscale framework that explicitly couples atomistic charge redistribution with heterogeneous stress fields associated with dislocation networks. In particular, integrating electronic charge-transfer models with continuum descriptions of dislocations would enable quantitative prediction of charge disproportion in complex defect configurations beyond isolated edge dislocations. Further extensions should systematically incorporate spin polarization and finite-temperature effects, especially in Cr- and Mn-containing alloys, to clarify the interplay between magnetic fluctuations \cite{moitzi2025inversion}, electron work function variations, and chemical softness under strain. In addition, the role of more complex defect topologies, such as dislocation junctions, stacking fault networks, and grain boundaries, remains to be explored, where non-linear stress superposition may further amplify or suppress anomalous charge-transfer behavior. On the methodological side, combining first-principles-informed charge equilibration schemes with machine-learned interatomic potentials \cite{anand2023exploiting} offers a promising route to scale these insights to larger systems while retaining electronic accuracy. Finally, direct experimental validation using local electrostatic potential probes, such as Kelvin probe force microscopy or electron energy-loss spectroscopy, would be valuable for establishing quantitative links between predicted electron work function variations and charge disproportion in concentrated alloys.

\section*{Acknowledgements}
 This research used resources of the Oak Ridge Leadership Computing Facility, a DOE Office of Science User Facility operated by the Oak Ridge National Laboratory under contract DE-AC05-00OR22725. 
This manuscript has been authored in part by UT-Battelle, LLC, under contract DE-AC05- 00OR22725 with the US Department of Energy (DOE). The publisher acknowledges the US government license to provide public access under the DOE Public Access Plan (http://energy.gov/downloads/doe-public-access-plan).

During the preparation of this work the authors used OpenAI (GPT-5) in order to search literature, correct grammar and revise language. After using this tool/service, the authors reviewed and edited the content as needed and takes full responsibility for the content of the published article.

\section*{Author contributions}
Gautam Anand:  Conceptualization,  Methodology,  Software, Formal analysis,  Investigation, Writing - Original Draft,  Writing - Review \& Editing,
Swarnava Ghosh:  Conceptualization,  Methodology,  Software, Formal analysis,  Investigation,  Writing - Original Draft,  Writing - Review \& Editing, 
Suman Chabri: Conceptualization, Methodology \& Investigation,
Markus Eisenbach:  Conceptualization,  Methodology,  Software, Formal analysis,  Investigation, Writing - Review \& Editing;

\section*{Conflict of interest declaration}
The author declares no competing interests.

\bibliographystyle{unsrt}
\bibliography{dislocation_HEA}

\end{document}